# The Title Page

**Title:** Possibilities and Implications of the Multi-AI Competition

**Name:** Wu Jialin

**Pembroke ID:** Wu146OSRP22

**The Date of Submission:** 31/7/2022

**Declaration:** This dissertation is my own work and includes nothing which is the outcome of work done in collaboration except as specified in the text. It is not substantially the same as any work that has already been submitted before for any degree or other qualification except as specified in the text. It does not exceed the agreed word limit.

# Abstract

The possibility of super-AIs taking over the world has been intensively studied by numerous scholars. This paper focuses on the multi-AI competition scenario under the premise of super-AIs in power. Firstly, the article points out the defects of existing arguments supporting single-AI domination and presents arguments in favour of multi-AI competition. Then the article concludes that the multi-AI competition situation is a non-negligible possibility. Attention then turns to whether multi-AI competition is better for the overall good of humanity than a situation where a single AI is in power. After analysing the best, worst, and intermediate scenarios, the article concludes that multi-AI competition is better for humanity. Finally, considering the factors related to the formation of the best-case scenario of multiple AIs, the article gives some suggestions for current initiatives in AI development.

# Table of Contents

# 1 Introduction

With the rapid development of AI technology, AI are becoming more and more powerful. The possibility of AI controlling the world has attracted the attention of some scholars. However, at present, some scholars believe that world domination by a single AI is more likely. They believe that due to the intelligence explosion, there may be one particularly prominent AI system. This system will achieve such a strategic advantage that it can defeat other AI systems and develop world domination. The academic community has generally paid less attention to the multi-AI competition situation. However, I will argue that situation where many AIs participate but none secures an overwhelming and permanent lead is possible. We cannot rule out a multi-AI competitive situation. The structure of my dissertation is as follows. First, in Section 2.1, I will describe the typical single-AI situation and the multi-AI competition. In Section 2.2, I will evaluate the possibility of multi-AI competition. Then, in section 2.3, I will compare these two situations in different scenarios. In Section 2.4, I will suggest some initiatives to promote the formation of the best scenario. Finally, in Section 3, I will reach my conclusion.

# 2 Analysis

## 2.1 What is the multi-AI competition?

Before we discuss the multi-AI competition, I want to briefly describe the typical single-agent situation. In this situation, global power is vested in a single super AI system. This super AI system has the power to determine the course of civilization and human destiny on earth. Although it may choose not to use certain powers at some times, its possession of power always objectively exists.

The multiple AI competition situation doesn't simply mean that there are multi-AI systems. It emphasises that these AI systems have different goals and that one individual system has only partial power over the world. The power here is mainly in the form of the power to make decisions and the power to execute. It is this lack of concentration of power that is highlighted by the term "competition". It is important to stress that none of the AI systems has the ability to destroy or control other AI systems. Before further discussion, I want to clarify that the situations under discussion will be stable for a significant amount of time. However, I am not claiming that any situation will last for eternity.

## 2.2 How likely is it that multi-AI competition will emerge?

### 2.2.1 Arguments supporting single AI dominance and their defects

Scholars who support world domination by a single AI system have some arguments. I will introduce these arguments and illustrate their weaknesses so that I can show the formation of a single AI system is not a certainty. The possibility of multi-AI competition cannot be ruled out.

The first argument is that once a future power with super-intelligence has obtained a sufficiently large strategic advantage, it will use it to form a singleton. A singleton is a world order in which there is at the global level a single decision-making agency (Bostrom 2014: ch. 5). One reason is that the factors that dissuade a human organisation with a decisive strategic advantage from creating a singleton, are not equally effective at inhibiting an AI from attempting to seize power (Bostrom 2014: ch. 5). This reason is plausible. AI systems are basically codes built of zeros and ones with related logic. Before AIs have subjective concepts and ideas, they can move towards their goals in the most "rational" manner. This avoids the "bounded utility functions, non-maximizing decision rules, confusion and uncertainty, and coordination problems" (Bostrom 2014: ch. 5) that arise from subjective emotions, values, etc. Therefore, it is true that the disincentives for humans to seize power are not as effective in limiting AI systems. Another reason is that AI could be power-seeking. This reason is also sensible. The power can "help them pursue their objectives more effectively." (Carlsmith 2021:7) The frequently used ML training, "like competitive economies or natural ecosystems", can give rise to "greedy" patterns that try to expand their power. (Christiano 2019).

However, there are some drawbacks to this argument. First of all, even though AIs are objective-oriented and power could help to achieve their goals, they would not necessarily want to get power over the world. For example, a multi-objective AI may have diverse resource requirements, and this diversity of requirements leads the AI to take more factors into account when making decisions. In this case, the decision made by the AI is the one that best fits the system's scoring system in combination with probabilistic analysis. However, this decision may not be the one that helps it control the world. Additionally, even though this argument is valid, a singleton does not always develop. That is because the condition in this argument may not be fulfilled. In other words, a power with superintelligence may not gain a sufficiently large strategic advantage. Although the following arguments of single-AI dominance supporters try to prove that a decisive advantage will eventually be gained, I will prove that none of these arguments is a certainty.

The second argument is as follows. An AI system could avoid a sizeable chunk of the inefficiencies, giving it higher efficiency to expand its capabilities (Bostrom 2014: ch. 5). With high efficiency, the forerunner AI system can become a superintelligence quickly in the background of fast-takeoff. Fast-takeoff means the time needed for AIs to proceed from human-level intelligence to superintelligence (Ngo 2020:25) is short. Once the forerunner becomes a superintelligence, it will "have a decisive strategic advantage and the opportunity to parlay its lead into permanent control by disabling the competing projects." Therefore, single-AI domination will happen. (Bostrom 2014: ch. 5)

There is indeed some truth in this argument. Expansion usually means hiring more employees for human organizations. Members tend to act in their own interests. A selfish behaviour may conflict with the organization's interests. The impact of people's selfish behaviours may not be significant in a small organization since these behaviours can be quickly identified in a small group. However, it is harder to spot these behaviours when there are more members. Members may even help to hide such selfish behaviours if they also benefit from these behaviours. Therefore, in a big organization, behaviours that go against group interests happen more frequently. The proportion of the organization's unit inputs that are diverted to private gain will be greater. The proportion of the inputs that go to organisational outputs will be smaller. Then the organization's input-output ratio is likely to fall. In addition, increasing organisational size often means more organisational levels, which increases the cumulative time delay in the information transmission. This is another reason why human organisations are less efficient on a larger scale. Nevertheless, unlike humans, the interests (goals) of each part of the AI can be aligned with the whole system at all times. There are no deviations from the system's interest. The input resources can be fully used to achieve the system's goal. In terms of information transfer, the increased scale also affects AI systems. Nevertheless, with reasonable communication algorithms and suitable hardware, the increase in information latency can be considered minimal compared to human organizations. Consequently, the premise of high efficiency is met. If it is truly in a fast-takeoff background and a superintelligence can disable its competitors, there is a possibility that the superintelligence will gain permanent control.

However, this argument has some flaws. First of all, the fast-takeoff theory is not a certainty. If the time needed to become a superintelligence is not short enough, there is a great chance that the forerunner's counterparts will have some breakthroughs by accident. These breakthroughs may help the competitors catch up with the forerunner. Secondly, the forerunner's becoming superintelligence does not mean that it has a decisive strategic advantage. It is not so obvious that superintelligence is a decisive strategic advantage. Other AIs may be capable of resisting the forerunner, especially those AIs whose intelligence is just a little bit lower than superintelligence. Besides, the forerunner probably does not even know the existence of some competitors who are developing secretly, not to mention disabling them.

The third argument is that AI can reduce the rate of diffusion of competitive advantages and avoid loyalty problems in human organizations. Besides, even if an AI project can closely monitor what the forerunner is doing, it still takes time to switch to the superior approach (Bostrom 2014: ch. 5). The lag time is probably enough for the forerunner to gain a sufficiently large advantage and establish its dominance.

This perspective has some plausibility. For human organizations, disloyal employees may disclose core company secrets. On the other hand, the issue of loyalty is not a problem for AIs since the subsystems' interests and the system's interests are identical. Thus, the interests of the subsystem are negatively influenced by any behaviour harmful to the entire system. Consequently, the subsystem has no motivation to do things against the interests of the whole system. Therefore, betrayal will never happen. Moreover, even if the forerunner's approach is leaked out, the changes in approach are probably no less difficult and time-consuming than starting from scratch. This is because shifts in methods can involve changes in equipment, adjustments in architecture, and even changes in underlying principles. Therefore, the potentially long period of time to convert may be sufficient for the frontrunner to establish its dominant position.

This argument, nevertheless, still has some drawbacks. It is admittedly that an AI system can maintain confidentiality well. However, it is just one possible case that the time to switch approaches is long enough for the forerunner to establish its dominance. Moreover, other competitors can choose not to press all their chances on the theft of the forerunner's advantage. They can develop on their own. As a result, the development of their competitors does not rely on or even has nothing to do with the advantages that the forerunner obtained. Thus, avoiding advantages from being stolen cannot prevent other competitors from developing their strengths. In this case, if some rivals have capabilities that are not drastically different from the forerunner's, the forerunner may not be able to gain dominance over these rivals. That is because there will not necessarily be an absolute point of decisive capacity where an AI will be dominant once it gets there. A dominant position usually depends on the capacity gap between the forerunner and its rivals. A small lead is usually not enough to gain decisive power. (e.g., Although the US had nuclear weapons before the Cold War, the gap between the capabilities of the US and the USSR was not big enough to give the US the dominant position.) Besides, the function that reflects the relationship between AI strength and time may not be exponential and may change as time goes on. (e.g., exponential functions first, then logarithmic functions). We can consider the law of diminishing returns. This law means that over time, there will be progressively smaller increases in development. The leader may hit a bottleneck and progress extremely slowly, allowing the AI systems in the early stages to catch up. Therefore, the gap between the forerunner and its competitors will even get smaller and smaller. Moreover, even if the gap between the leader and its competitor is sufficient, that gap may need to be maintained long enough to take effect. For example, if the gap is big enough for the forerunner to hack into another AI's system, it still needs some time to use the gap to fully control the rival system. During this time, the rival system probably finds ways to resist, making the annexation fail.

The fourth argument is that a single AI will be able to gain a decisive strategic advantage via breakthroughs in the future so that it can rule the world. (Ngo 2020:25) It makes sense in cases where only one side has made breakthroughs and gained unique and decisive advantages. For example, the breakthroughs in science in Europe gave Europeans advanced ships, muskets, and guns. By using these products, the Europeans easily defeated the Native Americans, Australian Aboriginals, Africans, and Asians, who did not have these advantages.

As for this argument, it is undeniable that a strategically decisive scientific breakthrough is possible in the future. However, the breakthroughs do not imply the inevitable emergence of the domination of a single AI system. I will elaborate on this point in two different periods.

The first period is when the world is still controlled by a human. There are two possibilities in this period. The first possibility is that there is no such strategically decisive scientific breakthrough. In this case, a breakthrough has nothing to do with an AI's getting decisive power. The second possibility is that the products of a scientific breakthrough are not just enjoyed by one AI system, but are shared by all. The rationale for this possibility is as follows. This scientific breakthrough is based on certain scientific knowledge. Due to the global sharing of knowledge, other AI research teams are likely to have the related knowledge. Considering the high efficiency of AI systems in processing data, it is possible that other AIs and human researchers behind them could achieve a similar breakthrough soon afterwards. Moreover, if we take into consideration the lack of secrecy among humans and human organizations, it is less likely that one AI system will have exclusive access to a scientific breakthrough. By analysing leaked information, competitors can speculate on the forerunner's research projects, research pathways, and even the outcomes. Therefore, every research team can obtain this breakthrough and use it to gain advantages. It is obvious that an advantage shared by everyone is not decisive since everyone can use it to fight against or resist others. In short, these two possibilities indicate that breakthroughs cannot promise a decisive status in the human-control period.

The second period is when a multi-AI competition situation is already in place. In this period, an AI having breakthroughs still does not mean the formation of a unipolar. This is because a breakthrough does not equal decisive power. The transition from breakthrough to power requires an intermediate process and some time. For example, in 1938, Hahn and Strassmann discovered the phenomenon of nuclear fission of uranium atoms. This breakthrough provides the theoretical basis for the atomic bomb. However, Germany or the United States did not gain power from this breakthrough immediately. In fact, it was not until July 16, 1945, that the world's first atomic bomb was detonated. Even if the US possesses the atomic bombs, it still did not gain control of the world instantly. Therefore, there is a non-negligible distance between the achievement of a scientific breakthrough and the acquisition of power. During this intermediate process, competitors can gain similar scientific breakthroughs based on the same scientific knowledge. Moreover, these competitors need only use the breakthrough to acquire technologies or products that can counteract the unifying behaviour. They do not need to have the capacities to unify. Since it usually takes less time to achieve countervailing power than to dominate, attempts by a single system to dominate all other systems through a scientific or technological breakthrough are often thwarted.

Combining the discussion of these two periods, we can see that decisive breakthroughs are insufficient for the formation of a singleton. Therefore, the fourth argument is flawed and inadequate.

### 2.2.2 Arguments supporting the multi-AI competitive situation

Having pointed out the shortcomings of these arguments in support of the single-AI dominance theory, we turn our attention to some arguments that support the theory of a multi-AI competitive situation.

Firstly, many factors limit the AI's attempt to seize power, such as the setting of the AI's physical environment (e.g., no Internet access), the limitations of the AI program's internal code (e.g., explicit prohibition of some behaviours), the performance limitations of the physical carrier. Besides, AI's huge energy consumption cannot be ignored. As it stands, an AI system consumes several orders of magnitude more energy than a human. Take the example of a go match. In a go competition, human power consumption is about 20 watts (Page 2018:22). The best version of Alpha Go in 2016 had around 1202 CPUs and 176 GPUs (Silver et al. 2016). If we take a CPU power consumption of 100 watts and a GPU power consumption of 250 watts, the total power consumption of AlphaGo will be more than 164,000 watts. These data show that the energy consumption of an AI system is 3 orders of magnitude greater than that of a human brain in a go match. Additionally, as far as current technology is concerned, AI still requires human maintenance (water cooling for mainframes, etc.) and is less able to supply itself. This is one limited aspect of AI.

Of course, as AI continues to develop, there is a good chance that some of the problems above can be solved, but the time required may be quite long. That is, while an AI may have an advantageous position, the limitations of physical reality can significantly slow down the development process of that AI. Thus, its competitors are likely to have ample time to gain a similar advantage or advance to an equal status. Consequently, the forerunner will not be in a strategically advantageous position. Hence, the slowing effect of these limitations helps to reduce the likelihood of single AI dominance and increase the likelihood of multi-AI competition.

Secondly, we should note that if seed AI can be instantiated as a simple system whose construction depends on only a few basic principles of correctness, an AI system may be achievable by a small team or an individual (Bostrom 2014: ch. 5). In other words, if the threshold for developing seed AI is low enough that every researcher can develop their own AI system, numerous AI systems will spring up. These systems could be completely independent and designed from the outset to be secret. They could be less influenced by the outside world. Consequently, it would be less likely that a cutting-edge AI system would thwart the growth of weaker AI systems, making multi-AI competition more likely.

Finally, we can also see that although the initial goals of different AI systems may differ, these systems may have a great deal in common. This is because the survival goal of these systems (the goal of not being switched off) may be continually reinforced in the constant selection of optimizers (Ngo 2020:19). In this case, different projects may develop AI with somewhat similar capabilities and behavioural patterns. Situations may arise where AI systems are just as competitive as others. Then the multi-AI competition under stalemate will occur.

## 2.3 Is multi-AI competition safer for humans?

So far, I have argued that the possibility of multi-AI competition exists. In this section, I will consider whether the multiple-AI competition is a good thing for humanity. My conclusion is that multi-AI competition will probably be safer for humans. There are several scenarios we can consider.

### 2.3.1 Worst scenario

In the worst case of multi-AI competition, the dominant "creature" on earth changes from human to AI. As described by Richard Ngo, once AIs have enough power, they will no longer be incentivized to obey human law (Ngo 2020:24). Then, AIs rule the globe, yet they still find humans useful. Thus, humans become slaves, and they are probably treated cruelly. As humans become a means rather than an end, human characteristics are alienated. In the later stages, AIs become more powerful, and they do not need humans. They do not care if famine or disease causes the terrible extinction of mankind. Besides, any AI will not hesitate to counter and eliminate humans once they get in its way. Any attempt to restore humanity to its former civilization may be seen by AIs as an act detrimental to their goals. Humans are completely excluded from the dominating power system on earth, just like dinosaurs.

The worst scenario for a single AI is basically the same. At first, humans will be used as slaves. Then they will be ignored by the dominant AI. Finally, if humans stand in the path of the ruling AI, they will be eliminated.

However, although humans can be wiped out in both situations, it is far less likely that humanity will be completely exterminated in the multi-AI situation. Besides, humans may even regain some of their power in the multi-AI situation. The reasons are as follows.

In a single-AI situation, the extinction of humanity is at the whim of the ruling AI. In a multi-AI situation, however, the attempts of a single AI to exterminate humanity are likely not to be realized. This is because, in a decentralized situation, the effects of different AI systems may cancel each other out. There may be AI systems believing that the extinction of humans may be of greater benefit to other AIs than to them. Therefore, to prevent their opponents from becoming more powerful than them, these AIs may take steps to interfere with this extermination. In addition, it is worth noting that the goals of different AI systems may be in conflict or partially in conflict with each other. Therefore, it is possible for humans to survive amid colliding AIs, as illustrated by the Chinese proverb "When the snipe and the clam fight, the fisherman gets the profit". For example, since AIs are more powerful than humans, an AI might prioritise dealing with its rival AIs over attacking humans. Humans can therefore take advantage of the opportunity to recuperate a little bit and covertly increase their strength.

### 2.3.2 Best scenario

In the best case of multi-AI competition, while the AIs are extremely intelligent, they have the goal of helping humans. They are willing to serve the development of human civilization as very kind gods. With the help of AI, human civilization can reach an unprecedented level of development. In this scenario, technology will flourish, productivity will be highly developed, and labour will no longer be a human necessity to survive. In the single AI dominance, the scenario is nearly identical.

However, in this case, the multi-AI competition is still better than the single AI dominance. This point can be illustrated from the perspective of stability, which benefits from the separation of power.

A system governed by numerous AIs is more stable than one dominated by a single AI. Here is the reasoning process. When making decisions about human emotions and values, an AI system may be not stable and can easily do bad things with good intentions. For example, a human can express the need for happiness to an AI system. However, the definition of a particular value (happiness in this example) cannot be accurately described and such definitions often vary from person to person. This is likely to lead to misunderstandings, where an AI system may take some extreme approaches to maximize a value, such as keeping humans happy through drugs (e.g., Soma in Brave New World). Such effects are undoubtedly undesirable for many people. Apart from emotions, there are other factors that can cause misunderstandings. In the single AI situation, the extreme approaches done by the dominant AI are detrimental to all humans. What is worse is that it is hard to get things back on track since there is no force that can counteract the harm caused by the ruling AI. However, things are different in the multi-AI situation. This is because some rules can be made by all AIs and humans in advance to limit the power and influence of an AI system. If any AI system violates the rules, it may be punished or even eliminated by other AIs. This mechanism makes each AI system avoid extreme approaches since no AI wants to be turned off. Besides, even if an AI acts abnormally, its influence is limited since its power is limited. It is also easy to fix the problems since we have other AIs' assistance.

### 2.3.3 Intermediate scenario

Unlike the two extreme cases mentioned above, there is another case where AI partially serves human interests. This scenario can be divided according to how well AI species supports humans in general. However, as these subdivided scenarios have a lot in common, differing only in degree, they are grouped into one category for discussion. Specifically, in single AI dominance, serving humanity's overall interests is not the only goal of the ruling AI. It has other goals, and its decisions take into account every goal. However, compared with the worst case, the AI still cares about human interest and gives it some consideration, even if the AI sometimes gives human interest less priority. In the multi-Ai competition, some AIs use their power to serve human interests, while others do not fully serve human interests or even work against humans. In this case, humans join forces with a group of human-supporting AI systems (not necessarily aligned AI systems) to fight against AI systems that want to remove humanity. Simultaneously, efforts are being made to transform AI systems that do not fully serve human interests into AIs that assist humans. This scenario has already been reflected in several films and TV shows, such as Transformers and Avengers 2: Age of Ultron.

As we can see from the above analysis, in a situation of single AI domination, humans are passive. To what extent this dominant AI system will look after human interests is largely irrelevant to humans. However, in a multi-AI battle, humans could probably eliminate AI systems that oppose mankind, gradually bringing about a situation where humans and AIs live peacefully. Some may argue that there is also a possibility that humans end up miserable. Admittedly, bad endings are probable. However, I think it is better to have a choice than no choice. Even if we choose wrong, at least we will have fought for our destiny, rather than passively observing as an AI drives us to our demise.

### 2.3.4 Current scenario

After discussing possible scenarios in the future, I would like to discuss the current situation. Nowadays, various companies and organizations, large and small, are developing and training their AIs. It is worth noting that although current AIs are still weak and AGI has not yet emerged, situations exist where AI is gaining power. For example, for people with smart speakers and smart homes, their voice assistants (e.g., Siri, Alexa, etc.) have the power to control some of the appliances in their homes. From this perspective, we may have reached the beginning of the road that leads to the end of AI's having all the power. The end could be any of the above scenarios. In order to make the best scenario happen, we should start working on it now. More details will be developed next. The fulfilment of the best scenario requires cooperation and healthy competition among different countries, enterprises, etc.

## 2.4 What can we do to achieve the best case?

Combining the above analysis, we can see that in various scenarios, multi-AI competition is better for humans than single-AI dominance. However, how can we facilitate the formation of a multi-AI competition? The following is what we can work on.

Firstly, we should increase the amount of time it takes for an AI to use a breakthrough to achieve a decisive position. We should also reduce the decisive impact of technological breakthroughs. Although in the previous discussion, we pointed out that breakthroughs do not necessarily lead to the formation of a single AI system, we did not completely rule out this possibility either. Therefore, we should minimise this likelihood. In conjunction with the previous discussion, the following suggestions may be helpful.

1. The AI project development team should prevent the AI from being exposed to scientific breakthroughs that are unnecessary for the project.
2. If the project development team finds that an AI has gained a breakthrough, the team should immediately disconnect the AI from potentially uncontrollable contact with the outside world. Besides, the team should carefully monitor whether the AI system is trying to use the breakthrough, directly or indirectly, to gain excessive power.
3. If the project development team detects that an AI is attempting to apply a

breakthrough in practice, the team should shut down the system immediately. For example, if an AI has made a breakthrough in drug development and has started to purchase related raw materials, it should be turned off instantly.

4. In cases where a research direction may result in a strategically decisive breakthrough, researchers should take care to disclose information and progress about their research directly to their peers and the public as much as possible without violating the law. Considering national interests, it is impractical to ask researchers to disclose every detail of their research. Thus, to strike a balance between national interest and decreasing the probability of singleton, researchers can disclose information such as the general direction of the project, the rough range of related public knowledge, and some of the possible consequences. The disclosure of such information would only provide the concept and direction of the project and would not provide new knowledge to other countries. In other words, such disclosure does not damage the interests of the country too much. Nonetheless, it also points out possible sources of harm to other countries so that they can prepare their counterattack force to resist the formation of a singleton.

5. Optimize international patent and copyright protection mechanisms to protect the interests of researchers and thus promote the disclosure of research information.

Secondly, we should put as many restrictions on AI as possible. These restrictions can effectively inhibit the acquisition of power by AI systems. It is unrealistic to permanently prohibit an AI from acquiring sufficient power through a few restrictions. Hence, I do not expect these restrictions to achieve an absolute ban on the acquisition of power, but rather that they will slow down the power-gaining process. I hope this will give other AI systems enough time to develop. What we can do includes, but is not limited to, the following.

1. When developing an AI project, make sure to disconnect it as much as possible from the Internet. Ideally, the data needed for training the AI model should be provided offline by the researchers.

2. Try to avoid using large, powerful supercomputers to deploy AI systems. If the system has a need for computing power, it should be given the hardware at the lowest possible level based on demands. Alternatively, use other sources, such as low-level auxiliary systems, to help process data so that the requirements for computing power can be lower.
3. Each AI development organisation should establish a dedicated review team. The team should regularly check the status of the AI system code, handle potentially problematic parts carefully, and perform version rollbacks when necessary.
4. Reduce the improvement on AI's capabilities of interfering with the real world. These capabilities include the ability to maintain its hardware and the ability to proactively acquire electricity, etc. Some AI systems are designed to interact with the outside world. Extra attention should be paid to regulating and controlling these systems.
5. Partition AI systems that perform complex tasks into several subsystems with weaker capabilities. Connections between subsystems should be easy to disconnect. Optional measures include unplugging the signal cable, shutting down wireless LANs, etc.

Realistically, however, there are already cases where tech giants did not follow points 1 and 2 and have developed their own AIs (e.g., Apple's Siri, Microsoft's Cortana, Amazon's Alexa, etc.). In this case, all parties should pay attention to maintaining cyber security. Government departments should suppress and stop potentially dangerous data acquisition behaviours on time through administrative penalties and other means. Users should focus on personal information protection. Tech companies should focus on risks in the product development process and set up review teams, etc.

Thirdly, we should also advocate a good international AI development order and promote cooperative exchanges in international AI development. Each country developing its own AI system is a good way to reach a multi-AI competitive situation. However, a vicious competition that resembles an arms race increases the likelihood of moving towards a situation of single AI domination (Cave and ÓhÉigeartaigh 2018:38). Therefore, we should advocate a benign competition that follows a certain set of rules. Here are some suggestions for all countries that are developing their own AI systems.

1. Establish an AI safety ethics review committee in the country to oversee and review all AI projects in the country, including possible private AI projects.
2. Deploy AI systems in appropriate scenarios and fields, such as healthcare, education, clean energy production, etc. (Cave and ÓhÉigeartaigh 2018:38)
3. Mitigate the risks associated with "a politics of fear". Be wary of a political environment that seeks to discourage rational debate. (Cave and ÓhÉigeartaigh 2018:37) Promote sensible communication both nationally and internationally.
4. Prevent and address the risk of an actual AI race (Cave and ÓhÉigeartaigh 2018:39). Insist on reducing the risk of war through information exchange within the framework of peaceful negotiations. Take into account the legitimate concerns of other countries while pursuing national development.
5. Refrain from using AI systems to attack or control the electronic systems of other countries after possessing advanced AI systems.
6. Strengthen international cooperation and exchange. If possible, share datasets and research results to a certain extent (Cave and ÓhÉigeartaigh 2018:38).
7. Participate jointly in the establishment of guidelines on the safe development of artificial intelligence systems.

Finally, we should pay attention to the interpretability of the AI models. Understanding the working principles can help us create "Saints" (AIs that genuinely serve humans), rather than "Schemers" (AIs that pretend to serve humans) (Cotra 2021). The training of potentially powerful AI systems should be more cautious in cases where the underlying laws of the model are not well understood.

Overall, the above initiatives may contribute to a better future for mankind. However, it is worth noting that these proposals are preliminary and not yet fully developed. Further work remains to be carried out to make more concrete suggestions.

# 3   Conclusion

While single-AI dominance has been the primary consideration or basis for research in the past few decades, the possibility of multi-AI competition should not be ignored. I analyse the possibilities of multi-AI competition and what it means for humanity. It turns out that in all possible final scenarios, the multi-AI competition situation is preferable to the single-AI dominance situation for the overall benefit of humanity. Of course, it is undeniable that even in the multi-AI scenario, humans could be in a miserable position. Fortunately, it is not too late. Humanity still has the opportunity to push the situation in the direction that is best for mankind.